\documentclass[12pt]{article}

\newif\iffigs\figstrue

\usepackage{epsfig,latexsym}
\usepackage{amsmath}
\usepackage{color}
\usepackage{verbatim}
\usepackage{mathrsfs}
\usepackage{amssymb}

\DeclareMathAlphabet{\mathpzc}{OT1}{pzc}{m}{it}

 \csname
@addtoreset\endcsname{equation}{section}

\def\gz0{\gamma^{0}}

\def\beq{\begin{equation}}
\newcommand{\eeq}[1]{\label{#1}\end{equation}}
\def\bea{\begin{eqnarray}}
\newcommand{\eea}[1]{\label{#1}\end{eqnarray}}
\def\ba{\begin{array}}
\def\ea{\end{array}}
\def\bec{\begin{center}}
\def\ec{\end{center}}
\def\ba{\begin{align}}
\def\ena{\end{align}}

\def\12{\frac{1}{2}}

\newcommand{\I}{\mathrm{Im}}

\usepackage{slashed}

\usepackage{float}

\usepackage{booktabs}
\usepackage{caption}

\usepackage[mathscr]{euscript}

\usepackage{color}
\usepackage{graphicx}
\usepackage{epsf}
\usepackage{graphicx,epsfig}
\usepackage{amsmath}

\usepackage{multirow}

\usepackage{amsmath, amsthm, amssymb}

\usepackage{epsfig}
\usepackage{cite}
\usepackage{color,colordvi}

\newcounter{hran}

\makeatletter
\renewcommand\section{\@startsection {section}{1}{\z@}%
                               {-3.5ex \@plus -1ex \@minus -.2ex}%
                               {2.3ex \@plus.2ex}%
                               {\normalfont\large\bfseries}}
\makeatother

\vspace{0.5cm}

\newcommand{\bi}{\begin{itemize}}
\newcommand{\ei}{\end{itemize}}

\begin{document}

\begin{flushright}
CERN-PH-TH/2015-017\\
\end{flushright}

\vspace{15pt}

\begin{center}


{\Large\sc Massive Born--Infeld and Other Dual Pairs}\\


\vspace{35pt}
{\sc S.~Ferrara${}^{\; a,\,b,\ *}$ \ and \  A.~Sagnotti${}^{\; a,\,c}$}\\[15pt]

{${}^a$\sl\small Department of Physics, CERN Theory Division\\
CH - 1211 Geneva 23, SWITZERLAND \\ }
\vspace{6pt}

{${}^b$\sl\small INFN - Laboratori Nazionali di Frascati \\
Via Enrico Fermi 40, I-00044 Frascati, ITALY}\vspace{6pt}

{${}^c$\sl\small
Scuola Normale Superiore and INFN\\
Piazza dei Cavalieri \ 7\\I-56126 Pisa \ ITALY \\}
e-mail: {\small \it sagnotti@sns.it}

\vspace{8pt}

\vspace{35pt} {\sc\large Abstract}
\end{center}
We consider \emph{massive} dual pairs of $p$--forms and $(D-p-1)$--forms described by non--linear Lagrangians, where non--linear curvature terms in one theory translate into non--linear mass--like terms in the dual theory. In particular, for $D=2p$ and $p$ even the two non--linear structures coincide when the non--linear massless theory is self--dual.  This state of affairs finds a natural realization in the four--dimensional massive $N=1$ supersymmetric Born--Infeld action, which describes either a massive vector multiplet or a massive linear (tensor) multiplet with a Born--Infeld mass--like term. These systems should play a role for the massive gravitino multiplet obtained from a partial super--Higgs in $N=2$ Supergravity.
\baselineskip=20pt
\vfill
\line(1,0){250}\\
{\footnotesize {$^*$On leave of absence from Department of Physics and Astronomy, U.C.L.A., Los Angeles CA
USA}}

\noindent

\setcounter{page}{1}

\pagebreak

\newpage
\section{Introduction}\label{sec:intro}

Non--linear theories of $p$--form gauge fields naturally arise in Superstrings and are clearly relevant in the description of brane dynamics as models for the partial breaking of Supersymmetry. The latter incarnation also presents itself in rigid theories with extended Supersymmetry. A prototype of these non--linear theories is the four--dimensional Born--Infeld action \cite{BI}, which in the bosonic case possesses the remarkable property of being electric--magnetic self--dual.

The $N=1$ supersymmetric completion of the Born--Infeld action \cite{deser,CF} has the further remarkable property of being the Goldstone action for $N=2$ Supersymmetry partially broken to $N=1$ \cite{HLP,BG,RT,FGP}. Alternatively, this action is seen to arise from an $N=2$ two--derivative action, quadratic in the Maxwell field strength, where half of the $N=2$ vector multiplet, {\it i.e.} a chiral multiplet, has been integrated out. This fate can result from a scalar potential that admits a vacuum with unbroken $N=1$ Supersymmetry and gives a large mass to the chiral multiplet \cite{APT,RT,FPS,ADT}.

The supersymmetric Born--Infeld action inherits the property of its bosonic
counterpart, in that it is self--dual in Superspace \cite{BG,RT,BMZ,KT,FK}. It is the aim of this paper to explore the mass generation mechanism induced in these non--linear theories by some additional geometric couplings that are connected to the Stueckelberg mass generation for $p$--form gauge fields. They admit a dual formulation in terms of $(D-p-1)$--form gauge fields, where the dual mass term results from a Green--Schwarz coupling \cite{GS} of their $(D-p)$--form field strength to a $p$--form gauge field. Or, equivalently, of their $(D-p-1)$--form gauge field to a $(p+1)$--form field strength. As a result, the non--linear curvature interactions present in one formulation are dual to non--linear generalizations of the mass term present in the other.

A particularly interesting situation presents itself when $D=2p$ for $p$ even, where the non--linear system can be self--dual \cite{BMZ}, so that the non--linear interactions maintain their functional form. A notable realization of this phenomenon is provided by the massive Born--Infeld system in four dimensions, where the one--form formulation realizes a Stueckelberg--Higgs phenomenon for the massive vector. Its dual, instead, describes a massive antisymmetric tensor theory with a non--linear mass term given by the same Born--Infeld action, albeit with the vector field strength that becomes pure gauge and is absorbed by the antisymmetric tensor field. The latter realizes the anti--Higgs mechanism \cite{ADF,CFG}, whereby a massless antisymmetric tensor (dual to a scalar) becomes massive absorbing a massless vector.

The plan of this paper is as follows. In Section \ref{sec:massive} we describe massless and massive dualities in a general setting where the mass is induced by a Green--Schwarz coupling. In Section \ref{sec:self_even} we describe the notion of self--dual actions \cite{GZ,AFZ}, which is only available for $D=2p$ and $p$ even, a particular case being the massive Born--Infeld system. In Section \ref{sec:massive_selfdual} we turn to the peculiar self--dual gauge--field systems in odd dimensions, which are only possible for $D=2p+1$ and $p$ odd. These are simple extensions of the systems considered by Deser, Jackiw and Teitelboim \cite{DJT} and by Pilch, Townsend and van Nieuwenhuizen \cite{PTVN}, which are also providing mutually dual descriptions, up to boundary terms. In Section \ref{sec:superspace} we concentrate on $N=1$ supersymmetric theories in four dimensions, and briefly recall the superspace self--duality of the massless Born--Infeld action. In Section \ref{sec:massive_superspace} we move to the massive case in superspace, and to this end we introduce the relevant $N=1$ superspace Green--Schwarz term \cite{CFG,FKLP}. There are again two dual formulations. One provides the superspace generalization of a Stueckelberg action, a self--interacting massive vector multiplet \cite{F,VP} with a non--linear, rather than quadratic, curvature part. The other involves a self--interacting massive linear multiplet, whose non--linear mass term is determined by the same supersymmetric Born--Infeld action built up, originally, in terms of the vector multiplet field strength. Finally, Section \ref{sec:conclusion} contains our conclusions and some prospects for future developments, while the Appendix contains an explicit derivation of the dual massive Born--Infeld action.

\section{Massless and massive dualities}\label{sec:massive}

Let us begin by recalling that, in the language of forms and with a ``mostly positive'' signature, the free actions for massive $p$--forms $B_p$ take the universal form
\beq
{\cal L} \ = \ \frac{k^2}{2} \ H_{p+1}(B_p) \, \wedge \star\, H_{p+1}(B_p) \ + \ \frac{m^2}{2} \ B_p \, \wedge \star \, B_p \ ,
\eeq{1}
where the field strength and the gauge transformations, which are a symmetry for $m=0$, read
\beq
H_{p+1} \ = \ d B_p \ , \qquad \delta\, B_p \ = \ d \Lambda_{p-1} \ ,
\eeq{2}
and $k$ is a dimensionless coupling. In components eqs.~\eqref{1} and \eqref{2} translate into the familiar $p$--dependent expressions
\beq
{\cal L} \ = \ - \ \frac{k^2}{2\, (p+1)!} \ H_{\mu_1 \ldots \mu_{p+1}} \, H^{\mu_1 \ldots \mu_{p+1}} \ - \ \frac{m^2}{2\, p\,!} \ B_{\mu_1 \ldots \mu_{p}} \, B^{\mu_1 \ldots \mu_{p}} \ ,
\eeq{1a}
and
\beq
H_{\mu_1 \ldots \mu_{p+1}} \ = \ (p+1)\, \partial_{[\mu_1} B_{\mu_2 \ldots \mu_{p+1}]} \ , \qquad \delta\, B_{\mu_1 \ldots \mu_p} \ = \ p\, \partial_{[\mu_1} \Lambda_{\mu_2 \ldots \mu_{p}]} \ ,
\eeq{2a}
where the antisymmetrizations have strength one.

Let us also recall that, in $D$ dimensions and with the given ``mostly positive'' signature, for any $p$--form $B_p$
\beq
\star \star B_p \ = \ - \ (-1)^{p\,(D-1)} \ B_p \ .
\eeq{3}

Our starting point is quite familiar and plays a central role in Supergravity. It will serve to fix our notation, and concerns the standard $D$--dimensional duality between a \emph{massless} ``electric'' $p$--form $B_p$ and a corresponding ``magnetic'' $(D-p-2)$--form $A_{D-p-2}$. It involves the corresponding field strengths $H_{p+1}=d B_p$ and $F_{D-p-1}=d A_{D-p-2}$, and is encompassed by the ``master action''
\beq
{\cal L} \ = \ \frac{k^2}{2} \ H_{p+1} \, \wedge \star\, H_{p+1} \ + \
H_{p+1} \wedge d C_{D-p-2} \ .
\eeq{4}

In this first--order action $H_{p+1}$ starts out as an unconstrained field, and the massless version of eq.~\eqref{1} is recovered once one enforces the field equation of $C_{D-p-2}$. This is precisely the Bianchi identity for $H_{p+1}$, so that in flat space its solution is given in eq.~\eqref{2}. Conversely, integrating out $H_{p+1}$, which is unconstrained to begin with, as we have stressed, yields
\beq
k^2 \ \wedge \star\, H_{p+1} \ = \ - \ d C_{D-p-2} \ ,
\eeq{5}
Substituting in eq.~\eqref{4} and making use of eq.~\eqref{3} leads finally to the dual representation
\beq
{\cal L} \ = \ \frac{1}{2 \, k^2} \ d C_{D-p-2} \wedge \star \, d C_{D-p-2} \ ,
\eeq{6}
with an inverted overall coefficient.

Let us now move on to consider a pair of gauge forms, $B_p$ and $A_{D-p-1}$, with the corresponding field strengths $H_{p+1}=dB_p$ and $F_{D-p}=dA_{D-p-1}$, and let us begin by considering a Lagrangian of the type
\beq
{\cal L} \ = \ \frac{k^2}{2} \ H_{p+1}(B_p) \wedge \star \, H_{p+1}(B_p) \ + \
{\cal L}_2\left[ F_{D-p}(A_{D-p-1}) \right] \ .
\eeq{7}
This Lagrangian describes a pair of massless gauge fields, $B_p$ and $A_{D-p-1}$.
Let us also refrain, for the time being, from making any definite assumptions on the nature of the Lagrangian ${\cal L}_2$, but let us add to this action principle a Green--Schwarz coupling. This carries along a coefficient $m$ of mass dimension, and one is thus led to
\beq
{\cal L} \ = \ \frac{k^2}{2} \ H_{p+1}(B_p) \wedge \star \, H_{p+1}(B_p) \ + \ m \ H_{p+1}(B_p) \wedge A_{D-p-1}\ + \
{\cal L}_2\left[ F_{D-p}(A_{D-p-1}) \right] \ ,
\eeq{8}
which is gauge invariant up to a total derivative.

We can now show that this Lagrangian admits two dual forms, which can be reached turning the description of either of the two gauge fields $B_p$ or $A_{D-p-1}$ into a first--order form. In four dimensions, the duality that we are exploring concerns, for instance, a massive electromagnetic potential and a massive two--form, both of which carry three degrees of freedom. It is sharply different from the massless ones, which link vectors to vectors or two--forms to scalars.

Let us stress that in the \emph{massive} cases that we shall shortly reach from eq.~\eqref{7} the two fields $B_p$ and $A_{D-p-1}$ describe, of course, identical numbers of degrees of freedom. Moreover these identical numbers coincide with the sum of the degrees of freedom carried by corresponding \emph{massless} fields, a correspondence that reflects the well--known binomial identity
\beq
{\binom{D-2}{p}} \ + \ {\binom{D-2}{D-p-1}}  \ = \ {\binom{D-1}{p}} \ .
\eeq{9}
This is the counterpart, in the massive construction, of the binomial identity
\beq
{\binom{D-2}{p}}  \ = \ {\binom{D-2}{D-2-p}} \ ,
\eeq{9c}
which underlies the more familiar massless dualities.

The identity \eqref{9} reflects the fact that eq.~\eqref{8} is a Stueckelberg realization, a property that we shall shortly make manifest. Notice that a correspondence of this type plays a role in String Theory, in the behavior of the massive perturbative excitations. These modes are indeed described, in the light--cone formalism, as would pertain to massless ones, but they occur in combinations that build up an extra dimension, and thus their masses, in a Stueckelberg realization. This fact has long led to a widespread belief that the underlying phenomenon ought of take the form of a huge spontaneous breaking in an eventual more complete formulation.

Let us now move to a first--order form for $H_{p+1}$, replacing eq.~\eqref{8} with
\bea
{\cal L} &=& \frac{k^2}{2} \ H_{p+1} \wedge \star \, H_{p+1} \ + \ m \ H_{p+1} \wedge A_{D-p-1}\ + \ H_{p+1} \wedge dC_{D-p-2}   \nonumber \\  &&+ \
{\cal L}_2\left[ F_{D-p}(A_{D-p-1}) \right] \ ,
\eea{10}
where now $H_{p+1}$ is unconstrained. Notice that in the limit $m \to 0$ this Lagrangian describes, for $ k\neq 0$, a massless $p$--form and a massless $(D-p-1)$--form, while if also $k \to 0$ one is left with a single massless $(D-p-1)$--form. Finally, if $ k=0$ and $m \neq 0$ the Lagrangian becomes empty, since the field $A_{D-p-1}$ is then constrained to be pure gauge.

If one now integrates out $H_{p+1}$, the end result,
\bea
{\cal L} &=&  {\cal L}_2\left[ F_{D-p}(A_{D-p-1}) \right] \nonumber \\
&+& \frac{m^2}{2\,k^2} \ \left( A_{D-p-1} \ + \ \frac{1}{m} \ dC_{D-p-2} \right) \wedge \, \star \left( A_{D-p-1} \ + \ \frac{1}{m} \ dC_{D-p-2} \right) \ ,
\eea{11}
involves a Stueckelberg--like mass term for $A_{D-p-1}$.

Conversely, one can go to a first-order form for $F_{D-p}$, considering
\bea
{\cal L} &=& \frac{k^2}{2} \ H_{p+1}(B_p) \wedge \star \, H_{p+1}(B_p) \ - \ F_{D-p} \wedge \left[(-1)^{pD} \, m \, B_{p}\ - \ dE_{p-1} \right]   \nonumber \\  &&+ \
{\cal L}\left[ F_{D-p} \right] \ ,
\eea{12}
where we have performed a partial integration in the term depending on $m$. Integrating out $F_{D-p}$ now leads to the condition
\beq
{\cal C} \ \equiv \ m (-1)^{pD} \, B_{p}\ - \ dE_{p-1} \ - \ \frac{\partial_L {\cal L}_2}{\partial F_{D-p}} \ = 0 \ ,
\eeq{13}
where the derivative of ${\cal L}_2$ is a left derivative, while the resulting Lagrangian involves the corresponding Legendre transform of ${\cal L}_2$:
\beq
{\cal L} \ = \ \frac{k^2}{2} \ H_{p+1}(B_p) \wedge \star \, H_{p+1}(B_p) \ + \ {\cal L}_{\rm 2, \, dual} \left[ m (-1)^{pD} \, B_{p}\ - \ dE_{p-1} \right] \ ,
\eeq{12a}
where
\beq
{\cal L}_{\rm 2, \, dual} \left[ m (-1)^{pD} \, B_{p}\ - \ dE_{p-1} \right] \ = \ \left.\left\{\
{\cal L}\left[ F_{D-p} \right] \ - \ F_{D-p} \ \frac{\partial_L {\cal L}_2}{\partial F_{D-p}}\ \right\} \right|_{\cal C} \ .
\eeq{12b}
In particular, for $p=2$ this could describe a massive two--form with a Born--Infeld--like mass term if ${\cal L}_{\rm 2, \, dual}$ were the Born--Infeld action in $D$--dimensions. Note, however, that this action is not self--dual away from four dimensions, since in $D$ dimensions the massive dual of a two--form is a $(D-3)$--form.
\section{Self--dualities in even dimensions}\label{sec:self_even}

In four dimensions an even stronger instance of duality is possible. Taking $p=2$, the two fields at stake are a two--form $B_2$ and one--form $A_1$, and we would like to consider the case in which ${\cal L}_2$ is a Born--Infeld action.

Reverting to the conventional notation, our starting point is thus the master action
\bea
{\cal L} &=& \ - \frac{k^2}{12} \ H_{\mu\nu\rho}\, H^{\mu\nu\rho} \ - \ \frac{m}{4} \ \epsilon^{\mu\nu\rho\sigma} B_{\mu\nu}\, F_{\rho\sigma} \nonumber \\ &+& \frac{\mu^2}{8\, g^{\,2}} \ \left[ \,1 \ - \ \sqrt{1 \ + \ \frac{4}{\mu^2} \ F_{\mu\nu}\,F^{\mu\nu} \ - \ \frac{4}{\mu^4} \ \left( F_{\mu\nu}\,\widetilde{F}^{\mu\nu} \right)^2}\,\right]\ ,
\eea{13}
where we have introduced a dimensionless parameter $g$, the counterpart of the parameter $k$ that accompanies the two--form kinetic term. The parameter $\mu$ is the Born--Infeld scale factor, with mass--squared dimension, which sizes the non--linear corrections.

As above, a massive variant of the Born--Infeld action principle would obtain eliminating $H$ after moving to a first--order form where it is unconstrained. However, as we have seen, the additional field is just a standard Stueckelberg mode, so that for brevity we can just display the gauge--fixed Proca--like Lagrangian for the massive Born--Infeld vector,
\beq
{\cal L} \ = \ \ - \frac{m^2}{2\, k^2} \ A_{\mu}\, A^{\mu} \ + \ {\cal L}_{BI} \left(g,\mu, F_{\mu\nu} \right) \ ,
\eeq{14}
where
\beq
{\cal L}_{BI}\left(g,\mu, F \right) \ = \ \frac{\mu^2}{8\, g^{\,2}} \ \left[ \,1 \ - \ \sqrt{1 \ + \ \frac{4}{\mu^2} \ F_{\mu\nu}\,F^{\mu\nu} \ - \ \frac{4}{\mu^4} \ \left( F_{\mu\nu}\,\widetilde{F}^{\mu\nu} \right)^2}\,\right] \ .
\eeq{14a}

In the massless case, the self--duality of the Born--Infeld action would translate into the condition that
\beq
{\cal L}_{BI} \bigg(g\, ,\, \mu\, ,\, F_{\mu\nu}(A) \bigg)  \ = \ {\cal L}_{BI} \left(g^\prime=\frac{1}{g}\, , \, \mu^\prime = \frac{\mu}{g^{\,2}}\, , \, F_{\mu\nu}(C) \right) \ ,
\eeq{14b}
where $C$ is the dual gauge field. On the other hand, in the presence of the Green--Schwarz term ($m \neq 0$), one can eliminate the vector altogether and work, in the dual formulation, solely in terms of the two--form $B_{\mu\nu}$. The self--duality of the massless Born--Infeld theory then implies that the dual action involves a Born--Infeld mass term and reads
\bea
{\cal L} &=& - \ \frac{k^2}{12} \ H_{\mu\nu\rho}(B)\, H^{\mu\nu\rho}(B) \nonumber \\ &+& \frac{\mu^2}{8\,g^{\,2}} \ \left[ \,1 \ - \ \sqrt{1 \ + \ \frac{4\, m^2\,g^{\,4}}{\mu^2} \ B_{\mu\nu}\,B^{\mu\nu} \ - \ \frac{4\, m^4\, g^{\, 8}}{\mu^4} \ \left( B_{\mu\nu}\,\widetilde{B}^{\mu\nu} \right)^2}\,\right] \ .
\eea{15}
The massless limit can be recovered reintroducing the gauge invariant combination $m \, B  \ + \ dC$ before letting $m \to 0$. In this fashion the limiting Lagrangian describes a massless two--form, dual to a scalar, and a dual massless vector $C$.

The reason for the peculiar result in eq.~\eqref{15} is precisely that the Born--Infeld action possesses the property of reproducing itself under the Legendre transform of eq.~\eqref{13}, as summarized in eq.~\eqref{14b}. This is actually a special manifestation of a phenomenon that can occur whenever $p=D-p$, and therefore for $D=2p$ and $p$--forms $B_p$. In these cases, the field strength one starts from possesses the same index structure as the dual gauge field, and therefore one can contemplate the possibility of a self-reproduction as above. The corresponding condition would read
\beq
\left.{\cal L}\left[ g\, , \, \mu \, , \, F_{p} \right] \ - \ F_{p} \ \frac{\partial {\cal L}}{\partial F_{p}} \right|_{\frac{\partial_L {\cal L}}{\partial F_{p}}\, =\, m  \, B_{p}} \ = \ {\cal L}\left[\,g^\prime\, , \, \mu^\prime \, , \,  m\, B_p \, \right] \ .
\eeq{16}

This property of self--reproduction reflects, in fact, the invariance of the four--dimensional massless Maxwell system, or of generalizations thereof as the Born--Infeld system, under duality rotations, which translates into the condition \cite{GZ, KT, AFZ}
\beq
\epsilon^{\mu\nu\rho\sigma} \ \left[ 4\ \frac{\partial{\cal L}}{{\partial F^{\mu\nu}}} \ \frac{\partial{\cal L}}{\partial F^{\rho\sigma}}
\  + \ \frac{1}{g^{\,4}} \ F_{\mu\nu} \, F_{\rho\sigma} \right] \ = \ 0 \ .
\eeq{17}
The very form of this condition reflects the symplectic nature of four--dimensional duality relations, which finds a direct counterpart in all dimensions $D=2p$, with \emph{even} $p$ and forms $F_{\mu_1 \ldots \mu_{p}}$. In these cases one can indeed formulate analogs of the Born--Infeld system for $(p-1)$--forms $A_{\mu_1 \ldots \mu_{p-1}}$, with
\beq
{\cal L} \ = \ \frac{\mu^2}{8\, g^{\,2}} \ \left[ \,1 \ - \ \sqrt{1 \ + \ \frac{8}{p\,!\ \mu^2} \ F^2 \ - \ \frac{16}{(p\,!)^2 \ \mu^4 } \ \left( F \,\widetilde{F} \right)^2}\ \ \right] \ .
\eeq{18}

Let us stress that the invariance of actions like \eqref{18} under duality rotations implies that they are self--reproducing, in the massive case, under the transformation in eq.~\eqref{16}. Consequently, similar results hold, in the massive case, for the $2p$--dimensional duality between $p$--forms and $(p-1)$--forms with \emph{even} $p$. In these cases a non--linear completion of a kinetic term turns, after a duality transformation, into a non-linear completion of a mass term for the dual field with an identical functional form.

On the other hand, in even dimensions $D=2p$ with $p$ \emph{odd}, the product $F \,\widetilde{F}$ vanishes identically, and any Lagrangian ${\cal L}\left(F^2\right)$ is duality invariant, simply because dualities act as opposite rescalings on the self--dual and antiself--dual parts $F_{\pm}$, and moreover $F^2 \ \sim \ F_+\, F_-$. As a result, if the field strength is self--dual the Lagrangian reduces to a constant.

\section{Self--dual massive dualities in odd dimensions}\label{sec:massive_selfdual}

It is well known that in even dimensions $D=2p+2$, with $p$ even, one can impose \emph{real} self--duality conditions on massless $p$--forms, which halve their propagating degrees of freedom. These systems play an important role in Supergravity and in String Theory, both in six and in ten dimensions, as well as in the description of the string world--sheet, although it was recognized long ago that they do not admit conventional action principles \cite{MS}. In this section we would like to elaborate on their counterparts in the massive case.

To begin with, the very structure of eq.~\eqref{10} implies that, in the massive case, self--dual systems can only exist for $p$--forms in odd dimensions $D=2p+1$. The purpose of this section is to describe how the two types of action principles that have been associated to them in \cite{DJT} and \cite{PTVN} can be extended in general to describe dual pairs.

For a massive $p$--form $B_p$, the proper \emph{real} self--duality condition halving the corresponding degrees of freedom is
\beq
k^2 \ H_{p+1}(B_p) \ \equiv \ k^2 \ dB_p \ = \ m\, \star {B}_{p} \ ,
\eeq{19}
which is clearly possible only in \emph{odd} dimensions $D=2p+1$. Furthermore, the corresponding Lagrangians can only be formulated for \emph{odd} $p$. One possible form can indeed be deduced from eq.~\eqref{10}, and reads
\beq
{\cal L} \ = \ \frac{k^2}{2} \ H_{p+1}(B_{p})\wedge \star \, H_{p+1}(B_{p}) \ + \ \frac{m}{2} \ H_{p+1}(B_{p}) \wedge B_{p} \ .
\eeq{20}
From this expression one can see that, for even $p$, the second term would be a total derivative, so that one would be describing a massless $p$--form. On the other hand, for \emph{odd} $p$ eq.~\eqref{20} is a generalization of the classic three--dimensional result of Deser, Jackiw and Templeton \cite{DJT}.

Another formulation for massive self--dual fields was also proposed long ago by Pilch, Townsend and van Nieuwenhuizen \cite{PTVN}, for arbitrary $D=2p+1$. It describes the same number of propagating degrees of freedom, so that one would expect that a relation to the preceding one exist, although apparently this was not noticed in \cite{PTVN}.

One can exhibit this relation performing in eq.~\eqref{20} the Hubbard--Stratonovich transformation
\beq
{\cal L} \ = \ \frac{1}{2\, k^2} \ C_{p} \wedge \star \, C_{p} \ + \   H_{p+1}(B_p) \wedge C_p \ + \ \frac{m}{2} \ H_{p+1} \wedge B_p  \  .
\eeq{20p}
Integrating out $C_{p}$ one would simply recover eq.~\eqref{20}, while integrating out $B_p$ leads to the condition
\beq
B_p \ = \ - \ \frac{1}{m} \ C_p \ + \ d \Lambda \ ,
\eeq{21}
where the last term gives rise only to total derivatives in ${\cal L}$. Substituting in eq.~\eqref{20} then leads to the dual Lagrangian
\beq
{\cal L} \ = \ \frac{1}{2\, k^2} \ C_{p} \wedge \star \, C_{p} \ - \  \frac{1}{2m} \ C_{p} \wedge d C_p \ ,
\eeq{22}
up to $\Lambda$--dependent boundary terms. Notice that both coefficients are inverted, so that, like all preceding examples, this has the flavor of a strong--weak coupling duality, while the physical mass is in both cases
\beq
M_{phys} \ = \ \frac{m}{k^2} \ .
\eeq{23}
%

\section{$N=1$ Superspace formulation }\label{sec:superspace}

As we have seen, in four dimensions a massive vector is dual to a massive antisymmetric tensor, and the Stueckelberg mechanism for the former finds a counterpart in an ``anti--Higgs'' mechanism for the latter (in the sense that a massless $B_{\mu\nu}$, with one degree of freedom, eats a massless vector carrying two degrees of freedom).

This relation becomes particularly interesting in the supersymmetric context, since the supersymmetric extension of the Born--Infeld action is the Goldstone action for $N=2$ spontaneously broken to $N=1$. It thus describes the self interactions of an $N=1$ vector multiplet whose fermionic component, the gaugino, plays the role of Goldstone fermion for the broken Supersymmetry. When coupled to Supergravity, this system is expected to be a key ingredient in models for the $N=2 \to N=1$ super-Higgs effect of partially broken Supersymmetry. Consequently, the gaugino must be eaten by the gravitino of the broken Supersymmetry, which then becomes massive. In fact, because of the residual Supersymmetry the massive gravitino must complete an $N=1$ massive multiplet, which also contains two vectors and a spin--$\frac{1}{2}$ fermion \cite{FVN}. The Born--Infeld system, which contains in the rigid case a massless vector as partner of the Goldstone fermion, must therefore become massive. This provides in general a motivation to address massive Born--Infeld systems for $p$--forms.

In four--dimensional $N=1$ superspace the vector $A$ belongs to a real superfield $V$, while the two-form $B$ belongs to a spinor chiral multiplet $L_\alpha$ ($ \overline{D}_{\dot{\alpha}} \, L_\alpha = 0$). The two--form field strength $dA$ belongs to the chiral multiplet
\beq
W_\alpha(V) \ = \ \overline{D}^{\,2} D_\alpha V \ ,
\eeq{23}
which is invariant under the superspace gauge transformation
\beq
V \ \rightarrow \ V \ + \ \Lambda \ + \overline{\Lambda}  \qquad (\overline{D}_{\dot{\alpha}} \, \Lambda \ = \ 0) \ .
\eeq{23a}
This is the superspace counterpart of the familiar Maxwell gauge transformation $A \to A \ + \ d \lambda$. On the other hand, the three--form field strength $H=dB$ belongs to a linear multiplet $L$, which is related to $L_\alpha$ according to
\beq
L \ = \ i \left( D^\alpha\, L_\alpha \ - \ \overline{D}_{\dot{\alpha}}\, \overline{L}^{\,\dot{\alpha}}\right) \ ,
\eeq{24}
and satisfies the two constraints
\beq
D^2 L \ = \ \overline{D}^2 L \ = \ 0 \ .
\eeq{24a}
Note that $L$ is invariant under the gauge transformation
\beq
L_\alpha \ \rightarrow \ L_\alpha \ + \ W_\alpha(Z)\ ,
\eeq{25}
for any real superfield $Z$, due to the superspace identity
$D^\alpha \overline{D}^{\,2} D_\alpha = \overline{D}_{\dot{\alpha}} D^{\,2} \overline{D}^{\dot{\alpha}}$. Eq.~\eqref{25} is the superspace counterpart of the gauge transformation $B \to B \ + \ d z$ \footnote{Strictly speaking, the linear multiplet $L$ is the super field strength of the chiral multiplet $L_\alpha$. Only the latter ought to be called tensor multiplet, because it contains the tensor field $B_{\mu\nu}$. With a slight abuse of language, however, we use loosely the term linear multiplet for both.}.

In superfield language, the supersymmetric Born--Infeld action takes the form
\beq
{\cal L}_{BI}  \ = \  \left.{\cal F}\big[g\, , \,\mu\, , \, W^2(V)\, , \,\overline{W}^{\,2}(V)\big]\right|_D \  + \ \left. \frac{1}{2\, g^{\,2}} \ \left( W^2(V) \ + \ {\rm h.c.} \right) \right|_F\ ,
\eeq{26}
where ${\cal F}$ is given in \cite{CF,BG}. This expression clearly reduces to the supersymmetric Maxwell action for ${\cal F}=0$, and the leading non--linear order correction is clearly proportional to $\left. \frac{1}{\mu^{\,2}\,g^{\,2}}\ W^2(V)\,\overline{W}^{\,2}(V) \right|_D$.

Proceeding as in Section \ref{sec:massive}, it is now convenient to recast eq.~\eqref{26} in the first--order form
\beq
{\cal L}_{BI} \ = \ \left.\phantom{\frac{1}{2}} {\cal F}\big[g\,,\,\mu\, ,\, W^2,\overline{W}^{\,2}\big]\right|_D \ + \ \left. \frac{1}{2\,g^{\,2}} \ \left( W^2 \ + \ {\rm h.c.} \right) \right|_F
\ +  \left. \phantom{\frac{1}{2}} i \left( M^\alpha \, W_\alpha \ + \ h.c. \right)\right|_F
\ ,
\eeq{27}
introducing a dual potential $V_D$ and letting
\beq
M_\alpha = \overline{D}^{\,2}\, D_\alpha V_D \ .
\eeq{27}
Integrating out $V_D$ one recovers the original action \eqref{26}, while integrating out $W_\alpha$ yields the condition
\beq
\overline{D}^{\,2}\ \frac{\partial {\cal L}_{BI}}{\partial W^\alpha} \ \equiv \ \overline{D}^{\,2}\, \frac{\partial {\cal F}}{\partial W^\alpha} \ + \ \frac{1}{g^{\,2}} \ W_\alpha \ = \ - \ i \ M_\alpha \ .
 \eeq{28}
The equation of motion then follows from the Bianchi identity of the dual field strength $W_\alpha(V_D)$,
\beq
D^\alpha\, M_\alpha \ - \ \overline{D}_{\dot{\alpha}}\, \overline{M}^{\dot{\alpha}} \ = \ 0 \ .
\eeq{29}
As discussed in \cite{BG,KT}, the superspace Born--Infeld action enjoys a self--duality, which is the direct counterpart of what we have seen in components in eq.~\eqref{17} and translates into the condition \cite{KT}
\beq
\left. \phantom{\frac{1}{2}} \Im \left[ \ M^\alpha \, M_\alpha \ + \ \frac{1}{g^{\,4}} \ W^\alpha \, W_\alpha \ \right]\right|_F \ = \ 0 \ ,
\eeq{30}
where $\Im$ picks the imaginary part of the $F$--component.
Notice that this is trivially satisfied in $N=1$ Maxwell Electrodynamics, on account of the special form of eq.~\eqref{28} when ${\cal F}$ vanishes.
\section{Massive supersymmetric Born--Infeld and its superspace dual}\label{sec:massive_superspace}
We now turn to the supersymmetric version of the massive duality of Section \ref{sec:massive}. The counterpart of the Lagrangian \eqref{10} is
\beq
{\cal L} \ = \ \Phi(U) \ + \ L(U \ - \ m V) \ + \ {\cal L}_{BI} \left[g\,,\,\mu\, ,\, W_\alpha(V),  \overline{W}_{\dot{\alpha}}(V)\right] \ ,
\eeq{31}
where $L$ is a linear multiplet Lagrange multiplier and $U$ and $V$ are real superfields. The Lagrangian \eqref{31}, with the last term replaced by a standard quadratic super--Maxwell term $W^\alpha W_\alpha$, was considered in connection with $R+R^2$ theories in \cite{CFPS} and, more recently, for models of inflation, in \cite{FKLP}. $U$ is the superfield extension of an unconstrained $H_{\mu\nu\rho}$, and the presence of the arbitrary function $\Phi(U)$ reflects the freedom to dress the tensor kinetic term with an arbitrary function of the scalar field present in the linear multiplet.

Varying the action with respect to $L$ gives
\beq
U \ - \ m\, V \ = \ T \ + \ \overline{T} \ , \qquad \left(\overline{D}_{\dot{\alpha}} T \ = 0 \, \right)
\eeq{32}
and ${\cal L}$ becomes
\beq
{\cal L} \ = \ \Phi(T \ + \ \overline{T} \ + \ m\, V) \ +  \ {\cal L}_{BI} \left[ g\,,\,\mu\, ,\, W_\alpha(V),  \overline{W}_{\dot{\alpha}}(V)\right] \ .
\eeq{33}
This is the supersymmetric Stueckelberg representation of a massive vector multiplet. Making use of the gauge invariance of ${\cal L}_{BI}$, one can turn it into the Proca--like form
\beq
{\cal L} \ = \ \Phi(m\, V) \ +  \ {\cal L}_{BI} \left[ g\,,\,\mu\, ,\,W_\alpha(V),  \overline{W}_{\dot{\alpha}}(V)\right] \ ,
\eeq{33}
where $\Phi(mV)$ contains supersymmetric generalizations of vector and scalar mass terms, but also a scalar kinetic term. As we have stressed, the massive multiplet contains a physical scalar, which is the very reason for the presence of the arbitrary function $\Phi$. In particular, in this supersymmetric generalization of eqs.~\eqref{11} and \eqref{14}, the supersymmetric Proca--like mass term that extends $A_\mu^2$ is generally dressed by a scalar function.

The dual supersymmetric Born--Infeld action is the supersymmetric completion of eq.~\eqref{15}. It can be obtained integrating by parts the Green--Schwarz term in eq.~\eqref{31} and then going to a first--order form for $W_\alpha$, which requires the introduction of the dual gauge field $M_\alpha(V_D)$. To begin with, however, notice that one can integrate out $U$, thus replacing the first two terms with the Legendre transform
\beq
\psi(L) \ = \ \left. \left[\Phi(U) \ - \ U \, \Phi^\prime(U)\right] \right|_{\Phi^\prime(U) = - L} \ .
\eeq{34}
Combining all these terms, the first--order Lagrangian takes the form
\beq
{\cal L} \ = \ \left. \left\{\psi(L) \ + \ {\cal L}_{BI} \left[ g\,,\,\mu\, ,\,W_\alpha,  \overline{W}_{\dot{\alpha}}\right] \right\} \right|_D  \ + \ \left. \left\{i\, W^\alpha \left[ m L_\alpha \ + \ M_\alpha(V_D) \right] \ + \ {\rm h.c.} \right\} \right|_F \ ,
\eeq{35}
which is the supersymmetric extension of eq.~\eqref{12}. The notation is somewhat concise, since ${\cal L}_{BI}$ also contains an $F$--term, as we have seen in eq.~\eqref{26}.

Notice that, in this richer setting, $\psi(L)$ contains in general non--linear interactions of the massive scalar present in the massive linear multiplet, which is dual to the massive scalar of the massive vector multiplet. Both multiplets contain four bosonic degrees of freedom (a scalar and a tensor in the linear multiplet, and a scalar and a vector in the dual vector multiplet). In the massless limit the linear multiplet becomes dual to a chiral multiplet, while the vector multiplet becomes self--dual.

Integrating over $W_\alpha$ and using the self--duality of ${\cal L}_{BI}$ one finally gets
\beq
{\cal L} \ = \ \psi(L) \ + \ {\cal L}_{BI} \left[\frac{1}{g}\, , \, \frac{\mu}{g^{\,2}}\, , \, m \, L_\alpha \ + \ M_\alpha(V_D) \, ,\, m \, \overline{L}_{\dot{\alpha}} \ + \ \overline{M}_{\dot{\alpha}}(V_D) \right] \ ,
\eeq{36}
where both contributions contain a $D$--term and, as we have seen in eq.~\eqref{26}, ${\cal L}_{BI}$ also contains an $F$--term.

This implies, in particular, that the first--order non--linear correction to the linear multiplet mass term is proportional to $\left. \frac{m^{\,4}\,g^{\,6}}{\mu^{\,2}}\ L^\alpha\, L_\alpha\, \overline{L}_{\dot{\alpha}}\, \overline{L}^{\,\dot{\alpha}}\right|_D$.
Notice that for $m \neq 0$ $M_\alpha(V_D)$ can be shifted away, so that the superfield $L_\alpha$ acquires a Born--Infeld--like mass term, whose bosonic counterpart can be found in eq.~\eqref{14}.

\section{Discussion}\label{sec:conclusion}

In this article we have considered pairs of massive $p$--forms and $(D-p-1)$--forms and their dual descriptions, with emphasis on the peculiar features that can emerge for $D=2p$ and $p$ even. In these cases, if the massless dynamics enjoys a self--duality property, this leaves an imprint, in the massive theories, as a duality map between non--linear curvature terms in one formulation and non--linear non--derivative terms of the same form in the other. In the supersymmetric case the above phenomenon indicates that the massive linear multiplet dual to a Born--Infeld massive vector multiplet possesses, in general, non--derivative interactions that build up the very same Born--Infeld action with the field strength $W_\alpha$ replaced by the tensor multiplet $m L_\alpha$, but also non--linear curvature interactions encoded in a function $\psi(L)$. Systems of this type are expected to emerge as subsectors in the partial $N=2 \to N=1$ super--Higgs mechanism of extended Supergravity, where the massive gravitino must be accompanied by the other members of a massive multiplet with respect to the unbroken Supersymmetry. In four--dimensional $N=1$ Supersymmetry the gravitino multiplet contains in fact two massive vectors. One of them would be identified with the leftover portion of the $N=2$ Goldstone multiplet, whose fermionic part is eaten by the gravitino that becomes massive.

\subsection*{Acknowledgements} We are grateful to P.~Aschieri for stimulating discussions. A.~S. is on sabbatical leave, supported in part by Scuola Normale Superiore and by INFN (I.S. Stefi). The authors would like to thank the CERN Ph--Th Unit for the kind hospitality.
\begin{appendix}
\section{The dual massive Born--Infeld--like action}\label{sec:appendix}
\end{appendix}

Let us begin by considering the four--dimensional master Lagrangian
\bea
{\cal L} &=& \frac{\mu^2}{8\, g^{\,2}} \ \left[ \,1 \ - \ \sqrt{1 \ + \ \frac{4}{ \mu^2} \ F^2(A) \ - \ \frac{4}{\mu^4 } \ \left( F(A) \,\widetilde{F}(A) \right)^2}\ \ \right] \ - \ \frac{k^2}{12} \ H^2(B) \nonumber \\ &-& \frac{m}{4} \ \epsilon^{\,\alpha\beta\gamma\delta}\ F_{\alpha\beta}(A) \, B_{\gamma\delta}
\eea{a1}
involving a gauge field $A_{\mu}$ with the corresponding field strength $F_{\mu\nu}$, and a gauge field $B_{\mu\nu}$ with the corresponding field strength $H_{\mu\nu\rho}$. The Green--Schwarz coupling proportional to $m$ results in the introduction of a mass term for the vector, which emerges explicitly integrating it by parts and turning to a first--order form for $H$. One can then integrate out $H$, arriving finally at the Lagrangian of eq.~\eqref{14}.

Here we would like to outline the steps leading to the dual massive Lagrangian for $B_{\mu\nu}$ of eq.~\eqref{15}. The first step involves the transition to a first--order for $F$ and the introduction of two Lagrange multipliers $\lambda$ and $\sigma$, as in \cite{RT,BMZ,AFZ}. The first multiplier eliminates the square root, while the second reduces the term depending on $F \widetilde{F}$ to a quadratic expression. All in all, this turns eq.~\eqref{a1} into
\bea
{\cal L} &=& - \ \frac{k}{12} \ H^{\,2}(B) \ - \ \frac{\mu^{\,2}}{16\, g^{\,2}} \ \left( \sqrt{\lambda} \ - \ \frac{1}{\sqrt{\lambda}} \right)^2 \ - \ \frac{\lambda}{4\, g^{\,2}}\ F^2 \ + \  \frac{\lambda\ \sigma}{2\, \mu^{\,2}\, g^{\,2}} \ F \, \widetilde{F} \nonumber \\
&-& \frac{\lambda \, \sigma^{\,2}}{4\, \mu^{\,2}\, g^{\,2}} \ - \ \frac{m}{4} \ \epsilon^{\,\alpha\beta\gamma\delta}\ F_{\alpha\beta}\, B_{\gamma\delta} \ - \ \frac{1}{4} \ \epsilon^{\,\alpha\beta\gamma\delta}\ F_{\alpha\beta}\, \partial_\gamma\, C_{\delta} \ .
\eea{a2}
In the presence of $m$, the last term can be gauged away, and one can then integrate out $F_{\mu\nu}$, obtaining
\beq
F_{\mu\nu} \ = \ - \ \frac{m \, g^2}{\lambda} \ \frac{\widetilde{B}_{\mu\nu} \ - \ 2 \, \frac{\sigma}{\mu^{\,2}} \ B_{\mu\nu}}{\left( 1 \ + \ \frac{4\, \sigma^2}{\mu^4} \right)} \ .
\eeq{a3}
This result determines the three bilinears
\bea
F^{\,2} &=& - \ \frac{g^{\,4}\, m^2}{\lambda^{\,2}} \ \frac{\left( 1 \ - \ \frac{4\, \sigma^2}{\mu^4} \right)B^{\,2} \ + \ \frac{4\, \sigma}{\mu^2}\ B\,  \widetilde{B}}{\left( 1 \ + \ \frac{4\, \sigma^2}{\mu^4} \right)^2} \ , \\
F\, \widetilde{F} &=& - \ \frac{g^{\,4}\, m^2}{\lambda^{\,2}} \ \frac{\left( 1 \ - \ \frac{4\, \sigma^2}{\mu^4} \right)B\, \widetilde{B} \ - \ \frac{4\, \sigma}{\mu^2}\ B^2}{\left( 1 \ + \ \frac{4\, \sigma^2}{\mu^4} \right)^2}\ , \\
F \, \widetilde{B} &=&  \frac{g^{\,2}\, m}{\lambda} \ \frac{B^{\,2} \ + \ \frac{2\, \sigma}{\mu^2}\ B\,  \widetilde{B}}{\left( 1 \ + \ \frac{4\, \sigma^2}{\mu^4} \right)} \ ,
\eea{a4}
and the Lagrangian then takes the form
\beq
{\cal L} \ = \ - \ \frac{k}{12} \ H^{\,2}(B) \ - \ \frac{\mu^{\,2}}{16\, g^{\,2}} \ \left( \sqrt{\lambda} \ - \ \frac{1}{\sqrt{\lambda}} \right)^2 \ - \ \frac{\lambda \, \sigma^{\,2}}{4\, \mu^{\,2}\, g^{\,2}} \
- \ \frac{g^{\,2}\, m^{\,2}}{4\, \lambda} \ \frac{B^2 \ + \ \frac{2\,\sigma}{\mu^{\,2}} \ B \, \widetilde{B} }{\left( 1 \ + \ \frac{4\, \sigma^2}{\mu^4} \right)}  \ .
\eeq{a5}
The final result of eq.~\eqref{15} follows after integrating out $\lambda$ and $\sigma$.


\begin{thebibliography}{666}

\bibitem{BI}
M.~Born and L.~Infeld,
  Proc.\ Roy.\ Soc.\ Lond.\ A {\bf 144} (1934) 425.

\bibitem{deser}
S.~Deser and R.~Puzalowski,
  J.\ Phys.\ A {\bf 13} (1980) 2501.

\bibitem{CF}
S.~Cecotti and S.~Ferrara,
  Phys.\ Lett.\ B {\bf 187} (1987) 335.

\bibitem{HLP}
J.~Hughes and J.~Polchinski,
  Nucl.\ Phys.\ B {\bf 278} (1986) 147;
J.~Hughes, J.~Liu and J.~Polchinski,
  Phys.\ Lett.\ B {\bf 180} (1986) 370.

 \bibitem{BG}
J.~Bagger and A.~Galperin,
  Phys.\ Rev.\ D {\bf 55} (1997) 1091
  [hep-th/9608177].

\bibitem{RT}
M.~Rocek and A.~A.~Tseytlin,
  Phys.\ Rev.\ D {\bf 59} (1999) 106001
  [hep-th/9811232].

\bibitem{FGP}
S.~Ferrara, L.~Girardello and M.~Porrati,
  Phys.\ Lett.\ B {\bf 376} (1996) 275
  [hep-th/9512180].

\bibitem{APT}
I.~Antoniadis, H.~Partouche and T.~R.~Taylor,
  Phys.\ Lett.\ B {\bf 372} (1996) 83
  [hep-th/9512006].

\bibitem{FPS}
S.~Ferrara, M.~Porrati and A.~Sagnotti,
  JHEP {\bf 1412} (2014) 065
  [arXiv:1411.4954 [hep-th]];
S.~Ferrara, M.~Porrati, A.~Sagnotti, R.~Stora and A.~Yeranyan,
  arXiv:1412.3337 [hep-th], to appear in Fortschritte for Physik.

\bibitem{ADT}
L.~Andrianopoli, R.~D'Auria and M.~Trigiante,
  arXiv:1412.6786 [hep-th].

\bibitem{BMZ}
P.~Aschieri, D.~Brace, B.~Morariu and B.~Zumino,
  Nucl.\ Phys.\ B {\bf 574} (2000) 551
  [hep-th/9909021].

\bibitem{KT}
S.~M.~Kuzenko and S.~Theisen,
  JHEP {\bf 0003} (2000) 034
  [hep-th/0001068],
  PoS tmr {\bf 2000} (2000) 022.

\bibitem{FK}
  J.~Broedel, J.~J.~M.~Carrasco, S.~Ferrara, R.~Kallosh and R.~Roiban,
  Phys.\ Rev.\ D {\bf 85} (2012) 125036
  [arXiv:1202.0014 [hep-th]].

\bibitem{GS}
M.~B.~Green and J.~H.~Schwarz,
  Phys.\ Lett.\ B {\bf 149} (1984) 117.

\bibitem{ADF}
C.~Aragone, S.~Deser and S.~Ferrara,
  Class.\ Quant.\ Grav.\  {\bf 4} (1987) 1003.

\bibitem{CFG}
S.~Cecotti, S.~Ferrara and L.~Girardello,
  Nucl.\ Phys.\ B {\bf 294} (1987) 537.

\bibitem{GZ}
M.~K.~Gaillard and B.~Zumino,
  Nucl.\ Phys.\ B {\bf 193} (1981) 221.

\bibitem{AFZ}
P.~Aschieri, S.~Ferrara and B.~Zumino,
  Riv.\ Nuovo Cim.\  {\bf 31} (2008) 625
  [arXiv:0807.4039 [hep-th]].

\bibitem{DJT}
S.~Deser, R.~Jackiw and S.~Templeton,
  Phys.\ Rev.\ Lett.\  {\bf 48} (1982) 975.

\bibitem{PTVN}
  P.~K.~Townsend, K.~Pilch and P.~van Nieuwenhuizen,
  Phys.\ Lett.\ B {\bf 136} (1984) 38
   [Addendum-ibid.\ B {\bf 137} (1984) 443].

\bibitem{FKLP}
S.~Ferrara, R.~Kallosh, A.~Linde and M.~Porrati,
  Phys.\ Rev.\ D {\bf 88} (2013) 8,  085038
  [arXiv:1307.7696 [hep-th]].

\bibitem{F}
P.~Fayet,
  Nuovo Cim.\ A {\bf 31} (1976) 626.

\bibitem{VP}
A.~Van Proeyen,
  Nucl.\ Phys.\ B {\bf 162} (1980) 376.

\bibitem{FVN}
S.~Ferrara and P.~van Nieuwenhuizen,
  Phys.\ Lett.\ B {\bf 127} (1983) 70.

\bibitem{MS}
N.~Marcus and J.~H.~Schwarz,
  Phys.\ Lett.\ B {\bf 115} (1982) 111.

\bibitem{CFPS}
S.~Cecotti, S.~Ferrara, M.~Porrati and S.~Sabharwal,
  Nucl.\ Phys.\ B {\bf 306} (1988) 160.


\end{thebibliography}
\end{document}